\documentclass[3p,times,twocolumn]{myelsarticle}

\usepackage{amssymb} 
 
  \biboptions{sort,compress}

\usepackage[figuresright]{rotating} 
 
\usepackage{epsfig}

\pdfmapfile{+txfonts.map}

\newcommand{\eqn}[1]{eq.~(\ref{#1})}

 \newcommand{\beq}{\begin{equation}}
 \newcommand{\eeq}{\end{equation}}
 \newcommand{\be}{\begin{equation}}
 \newcommand{\ee}{\end{equation}}
 \newcommand{\beqa}{\begin{eqnarray}}
 \newcommand{\eeqa}{\end{eqnarray}}
 \newcommand{\bea}{\begin{eqnarray}}
 \newcommand{\eea}{\end{eqnarray}}

 \newcommand{\cL}{\mathcal{L}}
 \newcommand{\cG}{\mathcal{G}}
 \newcommand{\cO}{\mathcal{O}}

  \def\nnu{\nonumber}

\begin{document} 
 
\begin{frontmatter} 
 
 
 
 
\title{	
\vspace{-1cm}
 \rightline{\large IFT-UAM/CSIC-11-104}
 \rightline{\large FTUAM-11-67}
 \vspace{1cm}
Lepton flavor violation in minimal flavor violation 
extensions of the seesaw\tnoteref{$^*$}}

\tnotetext[$^*$]{Invited talk at the Workshop on \emph{${\rm e}^+{\rm
      e}^-$ Collisions from $\Phi$ to $\Psi$ (PHIPSI08)}, September
  19-22, 2011, Novosibirsk, Russia.  Based in part on work done in
  collaboration with R. Alonso, G. Isidori, L. Merlo and
  L. A. Mu\~noz.}

 
\author{Enrico Nardi}  

 \address{
INFN, Laboratori Nazionali di Frascati,
  C.P. 13, 100044 Frascati, Italy.\\
{\rm and:}  \\ 
Departamento de F\'{\i}sica Te\'orica,  
   C-XI, Facultad de Ciencias,   \\    
                Universidad Aut\'onoma de Madrid,       
                C.U. Cantoblanco, 28049 Madrid, Spain \\ 
{\rm and:}  \\ 
 Instituto de F\'{\i}sica Te\'orica UAM/CSIC, 
Nicolas Cabrera 15, C.U. Cantoblanco, 28049 Madrid, Spain
}

\begin{abstract} 
  A minimal flavour violation hypothesis for leptons can be
  implemented essentially in two ways that are compatible with a type-I seesaw
  structure with three heavy singlet neutrinos $N$, and  that satisfy the
  requirement of being predictive, in the sense that all lepton flavor
  violating (LFV) effects can be expressed in terms of low energy
  observables.  The first realization is $CP$ conserving and is based
  on the flavor group $SU(3)_e\times SU(3)_{\ell}\times O(3)_N$
  (being $e$ and $\ell$ the $SU(2)$ singlet and doublet leptons).  The
  second realization allows for $CP$ violation and is based on
  $SU(3)_e\times SU(3)_{\ell+N}$. I review the main features of the
  two schemes and point out their different implications for LFV
  observables.
\end{abstract} 
 
\begin{keyword} 
%
Neutrino Physics \sep Global Symmetries \sep Rare Decays 
%
 
\end{keyword} 
 
\end{frontmatter} 
 
 
 

\section{Introduction}
\label{sec:intro}
The assumption that the sources of breaking of the flavour symmetry
present in the standard model (SM) Lagrangian determine completely the
structure of flavour symmetry breaking also beyond the SM, is commonly
referred to as the Minimal Flavour Violation (MFV)
hypothesis~\cite{MFV,MFV2,DAmbrosio:2002ex}.  In the quark sector
there is a unique way to implement MFV: the two quark SM Yukawa
couplings are identified as the only relevant breaking terms of the
$SU(3)^3$ quark-flavour symmetry~\cite{DAmbrosio:2002ex}.  For the
lepton sector the same is not true: the SM cannot accommodate Lepton
Flavour Violation (LFV) because there is a single set of Yukawa
couplings (those of the charged leptons) that can always be brought
into diagonal form by rotating the three $SU(2)_L$-doublets
$\ell_\alpha$ and the three right-handed (RH) $SU(2)_L$-singlets
$e_\alpha$ ($\alpha=e,\,\mu,\,\tau$).
However, with the discovery of neutrino oscillation it has been
clearly established that lepton flavor is not conserved.  It is then
interesting to extend the MFV hypothesis to the lepton sector (MLFV)
by starting from a Lagrangian able to describe the observed LFV in
neutrino oscillations. The problem is that we do not know which
physics beyond the SM is responsible for these effects, and different
generalizations of the SM yield different formulations of the MLFV
hypothesis.

\section{Minimal effective theories for the seesaw}

A theoretically very appealing way to extend the SM to a dynamical
model that can account for strongly suppressed neutrino masses is the
type-I seesaw, where it is assumed that in addition to the SM leptons
($\ell$ and $e$) at high energies there is at least another set of
dynamical fields carrying lepton flavour: three SM singlets heavy
Majorana neutrinos $N_i$.  The gauge invariant kinetic terms for the
lepton fields $\ell_\alpha$, $e_\alpha$ and $N_i$ is:
\be
\label{eq:Kin}
\cL_{\rm Kin}=
\bar\ell_\alpha \not\!\!D_{\ell}\,\ell_\alpha+
\bar e_\alpha \not\!\!D_{e}\, e_\alpha +
\bar N_i \not\!\partial\, N_i\,,
\ee
where $D_{\ell}\,, D_{e}$ denote covariant derivatives.  The largest
group of flavour transformations that leaves $\cL_{\rm Kin}$ invariant
is $\cG = U(3)_\ell\times U(3)_N\times U(3)_e$.  We assume that $\cG$,
or some subgroup of $\cG$, is the relevant group of flavour
transformations, and we require that the only symmetry-breaking terms
can be identified with the parameters appearing in the seesaw
Lagrangian, that is:
\bea
- \cL_{\rm seesaw}&=& 
\nnu
\epsilon_e\, \bar \ell_\alpha Y_e^{\alpha\beta}e_\beta\,  H   
+ \epsilon_\nu\, \bar\ell_\alpha Y_\nu^{\alpha j}\,N_j\, \tilde H \\
&+&\frac{1}{2}\,\epsilon_\nu^2\,\mu_L\; \bar N^c_i\, Y_M^{ij}\, N_j
+{\rm h.c.}.   
\label{eq:seesaw}
\eea
The symmetry group can be decomposed as $\cG = U(1)_Y \times U(1)_L
\times U(1)_R \times \cG_F~$ where $U(1)_Y$ and $U(1)_L$ correspond to
hypercharge (that remains unbroken) and to total lepton number,
respectively; $U(1)_R$ can be identified either with $U(1)_e$ or with
$U(1)_N$, corresponding respectively to global phase rotations of $e$
or $N$, and
\be
\cG_F = SU(3)_\ell\times SU(3)_N\times SU(3)_e~, 
\ee
is the flavour group, broken at some large scale $\Lambda_F\gg\,$TeV.
Formal invariance of $\cL_{\rm seesaw}$ under $\cG_F$ is recovered by
promoting the Lagrangian parameters to spurions transforming as:
\begin{equation}
   \label{eq:generalspurions}
Y_\nu \sim (3,\bar 3,1);
\ \  \  
Y_M \sim (1,\bar 6,1);  
 \ \ \ 
Y_e \sim (3,1,\bar 3)\,. 
\end{equation}
As regards the two broken Abelian factors, $U(1)_L$ is broken (by two
units) by $\mu_L$, that is a spurion with dimension of a mass, while
$U(1)_R\,$ is broken by a dimension-less spurion $\epsilon_R\,$, where
$\epsilon_R$ denotes $\epsilon_e$ or $\epsilon_\nu$.

By itself, the Lagrangian~\eqn{eq:seesaw} induces LFV effects for the
charged leptons that are well below $\cO(10^{-50})$, and thus
unobservable. However, a theoretical prejudice states that {\it there
  is new physics at the {\rm TeV} scale}, since this is needed to
cure the SM naturalness problem.  It is then reasonable to assume that
at some scale $\Lambda_{NP} \ll \Lambda_F,\,\mu_L$, presumably around
or somewhat above the electroweak scale, other states carrying flavour
exist.  Integrating out these heavy degrees of freedom, as well as the
heavy RH neutrinos with masses $ \epsilon^2_\nu\mu_L > $ TeV, at $E
\ll {\rm TeV}$ we obtain an effective Lagrangian of the form:
\be 
\label{eq:LowE}
\cL_{\rm eff} = \cL_{\rm SM} + \cL^{\rm seesaw}_{D5}+
\frac{1}{\Lambda_{NP}^2} \sum_{i} c_i O_i^{(6)} + \ldots~.  
\ee
$\cL^{\rm seesaw}_{D5}$ is the Weinberg
operator~\cite{Weinberg:1979sa} that depends on the spurions (see
\eqn{eq:D5}).  $O_i^{(6)}$ denote generic dimensions-six operators
written in terms of the SM fields and of the spurions, and the dots
denote higher dimension operators.  Dimensions-six operators involving
only the SM fields conserve $B-L$~\cite{Weinberg:1979sa}, and since we
have not introduced (dangerous) sources of $B$ violation, then the
operators $O_i^{(6)}$ must conserve separately $L$.  This is the
reason why the scale $\Lambda_{NP}$ can be substantially lower than
$\Lambda_F$ and $\mu_L$.  Note also that $U(1)_N$ breaking and
$\epsilon_\nu$ only affect the RH neutrino masses, without affecting
in any way the Weinberg operator (see \eqn{eq:D5}).  As far as the
flavour structure of the operators $O_i^{(6)}$ is concerned, the
assumptions about $\cG_F$ breaking imply the following:
\begin{itemize}
\item[{\bf I.}] {\em Once the transformation properties of the
    spurions eq.}~(\ref{eq:generalspurions})~{\em and of the fields are
      taken into account},~{\em all $O_i^{(6)}$ must be formally
      invariant under $\cG_F$.}
\end{itemize}
This condition alone is not sufficient to obtain an effective theory
that is predictive, since the flavour structure of $Y_\nu,\,Y_M$ and
$Y_e$ cannot be determined from low-energy data
alone~\cite{Cirigliano:2005ck}.  A predictive MLFV formulation must
satisfy an additional working hypothesis:
\begin{itemize}
\item[{\bf II.}]  {\em The spurions flavour structure must be
    reconstructable from low energy observables, namely the light
    neutrino masses and the PMNS mixing matrix.}
\end{itemize}
The only way this second hypothesis can be satisfied is by restricting
the form of the spurions $Y_i$ in such a way that the relevant LFV
combinations will depend on a reduced number of parameters. This can
be obtained by assuming that the flavour symmetry corresponds to a
subgroup of $\cG_F$, rather than to the full flavour group.

\mathversion{bold}
\subsection{ $U(1)_R$ breaking  and size of the LFV effects}
\mathversion{normal}
Before analyzing the possible subgroups of $\cG_F$ yielding predictive
frameworks, let us discuss the connection between the overall size of
the LFV effects and the breaking of $U(1)_R$.  The explicit structure
of $\cL^{\rm seesaw}_{D5}$ and the corresponding
light neutrino mass matrix are
\bea
\label{eq:D5}
\cL^{\rm seesaw}_{D5} &=& \frac{1}{\mu_L}
\left(\bar\ell\tilde H\right) 
Y_\nu\frac{1}{Y_M}Y_\nu^T  \left(\tilde H^T\ell^c\right)\,, \\  
\label{eq:Mnu}
 \Longrightarrow\qquad 
  m_\nu^\dagger &=& 
\frac{v^2}{\mu_L}\; Y_\nu\frac{1}{Y_M}Y_\nu^T
\ = \ U\, \mathbf{m}_\nu\, U^T \,,    
\eea
where $v$ is the Higgs vacuum expectation value, $U$ is the PMNS matrix
and ${\mathbf m}_\nu = {\rm diag }
(m_{\nu_1},\,m_{\nu_2}\,,m_{\nu_3})$.  Note that since
$\cL^{\rm seesaw}_{D5}$ 
does not break $U(1)_{R}$, the overall size of 
${\mathbf  m}_\nu$ 
depends only on the lepton-number violating scale $\mu_L$,
but not on $\epsilon_{e,\nu}$. 
Without loss of generality we can rotate $Y_{e}$ and $Y_M$ 
to a diagonal basis. In terms of mass 
eigenvalues the diagonal entries can be written as:
\begin{equation}
  \label{eq:diageM}
(Y_e)_{\alpha\alpha}=
\frac{1}{\epsilon_e\,v}\, m_\alpha\,,  \qquad 
(Y_M)_{ii}=\frac{1}{\epsilon_\nu^2\,\mu_L}\,M_i \,. 
\end{equation}
This shows that the overall size of $Y_e$ and $Y_M$ is
controlled by the Abelian spurions (the same is true for $Y_\nu$).  A
natural choice for their size is such that the entries in the $Y_i$
matrices are of $\cO(1)$. Considering the light-neutrino mass
matrix \eqn{eq:Mnu} it can be seen how  
this choice points to a very large $L$-breaking
scale
\be
\label{eq:muL}
\mu_L \sim 
v^2/\sqrt{\Delta m^2_{\rm atm}} \approx  6\times 10^{14}~{\rm GeV}\,. 
\ee
In the case when $U(1)_R=U(1)_N$ however, we are free to assume
$\epsilon_\nu \ll 1$ as would naturally result from an approximate
$U(1)_N$ symmetry. In this case, in spite of the large values of
$\mu_L$, the RH neutrinos could have much smaller masses, possibly
within the reach of future experiments which, from the
phenomenological point of view, this represents a very interesting
possibility~\cite{Alonso:2011jd}.

\subsection{Two predictive cases}

The dimension-six LFV operators $O_i^{(6)}$ are invariant under
$U(1)_L$ and $U(1)_N$, but break   $\cG_F$ through 
various spurions combinations, like for example: 
\begin{equation}
  \label{eq:Delta}
\hspace{-.5cm}
  \Delta^{(1)}_8=Y_\nu Y_\nu^\dagger;\ \ 
  \Delta_6=Y_\nu Y_M^\dagger Y_\nu^T;\ \ 
  \Delta^{(2)}_8=Y_\nu Y_M^\dagger Y_M Y_\nu^\dagger\,. 
\end{equation}
In the absence of further assumptions, the $\Delta$'s cannot be
determined in terms of $U$ and ${\mathbf m}_\nu $.  To obtain
predictive frameworks basically two different criteria can be adopted,
that correspond to assume that in a given basis either $Y_M$ or
$Y_\nu$ are proportional to the identity matrix in flavour space
$I_{3\times 3}$~\cite{Cirigliano:2005ck,Alonso:2011jd}.  Both these
criteria have the property of being {\it natural} in the sense that
they can be formulated in terms of symmetry hypotheses, that is by
choosing as flavour symmetry some suitable subgroup of $\cG_F$.
(Alternative formulations of the MLFV hypothesis have also been
proposed in~\cite{Davidson:2006bd,Gavela:2009cd,Joshipura:2009gi}.)

\subsubsection{$SU(3)_N \to O(3)_N \times CP$.} 
Assuming that the flavour group acting on the RH neutrinos is $O(3)_N$
rather than $SU(3)_N$, implies that $Y_M$ must be proportional to
$I_{3\times 3}$. However, this condition alone is not enough to deduce
the structure of $Y_\nu$ from the seesaw formula. Full predictivity
for this framework is ensured only if we further assume that $Y_\nu$
is real: $Y_\nu^\dagger = Y_\nu^T$, which follows from imposing CP
invariance~\cite{Cirigliano:2005ck}. In this case, since the Majorana
mass term has a trivial structure, all LFV effects stem from the
(real) Yukawa coupling matrices giving:
\begin{equation}
  \label{eq:ON}
  \Delta_6 = \Delta_8^{(1)}=    \Delta_8^{(2)}= 
Y_\nu Y_\nu^T= \frac{\mu_L}{v^2}\; 
 U\, \mathbf{m}_\nu\, U^T \,. 
\end{equation}
The main implication for LFV in this scenario is that the largest
entries in the $\Delta$'s are determined by the {\it heaviest} neutrino
mass.  We refer to~\cite{Cirigliano:2005ck} for further details.

\subsubsection{\it $SU(3)_\ell\times  SU(3)_N \to SU(3)_{\ell+N}$.}
If we assume that $\ell$ and $N$ belong to the fundamental
representation of the same $SU(3)$ group, then in a generic basis
$Y_\nu$ must be a unitary matrix (and thus it can be always rotated to
the identity matrix by a suitable unitary transformation of the RH
neutrinos).  This condition, first proposed in~\cite{Alonso:2011jd},
also allows to invert the seesaw formula in \eqn{eq:Mnu}, giving
\begin{equation}
\Delta_6   
  =\frac{v^2}{\mu_L} \; 
  U\, \frac{1}{\mathbf{m}_\nu}\, U^T \,, \quad
  \label{eq:MMLFV2}
\Delta_8^{(2)}
  =\frac{v^4}{\mu_L^2} \; 
  U\, \frac{1}{\mathbf{m}_\nu^2}\, U^\dagger~,
\end{equation}
while $\Delta_8^{(1)} =I_{3\times 3}$ gives no LFV effects.  The
choice of a unitary $Y_\nu$ can be phenomenologically interesting
because it has been shown that if the $N$'s belong to an irreducible
representation of a non-Abelian group, then $Y_\nu$ is precisely
(proportional to) a unitary matrix~\cite{Bertuzzo:2009im}.  Now,
models based on non-Abelian (discrete) groups have proved to be quite
successful in reproducing the approximate tri-bimaximal~\cite{TBM}
structure of the PMNS matrix, so an approximate unitarity of $Y_\nu$
is what is obtained in several cases.  This scenario has also the
remarkable implication that the largest LFV effects are controlled by
the {\it lightest} neutrino mass. Other phenomenologically interesting
features are discussed in~\cite{Alonso:2011jd}.

Let us note at this point that $\mathbf{m}_\nu^{-1}$ appearing in
\eqn{eq:MMLFV2} does not correspond to any combination of the spurions
of the high energy theory. Therefore we learn that, contrary to common
belief, a MFV high energy Lagrangian can produce, at low energies,
operators which are not MFV.  A low energy theory following from a MFV
high energy theory is guaranteed to be also MFV only under the
additional requirement that, when all the spurions are set to zero,
the only massless fields are the SM ones.

\subsection{MLFV Operators}

Several MLFV operators can be constructed with the spurions
combinations given in \eqn{eq:Delta} or the analogous structures
involving also $Y_e$, like $ \Delta_{8}Y_e$, and it is useful to
provide at least a partial classification of the  most 
important ones: \\ [-6pt]

\noindent
{\it 1. On-shell photonic operators}.\\[2pt] 
They control the radiative decays $\ell\to \ell'\gamma$, 
and also contribute to $\mu-e$ conversion in nuclei, and  to 
four-leptons  processes like  $\ell\to 3 \ell'$ decays. Their structure is: 
\begin{equation}
  \label{eq:photonic}
O^{(F)}_{RL}  =  \bar\ell_\alpha
 \left(\Delta_{8}Y_e\right)^{\alpha\beta}
(\sigma\cdot F)\,  e_\beta\cdot H
\end{equation}
where $F$ denotes generically the field strength of the $SU(2)_L
\times U(1)_Y$ gauge fields. When these operators are the dominant
ones, one can predict quantitative relations between $\mu\to e\gamma$
and other processes, as for example:
\begin{eqnarray}
B_{\mu\to eee} &\simeq &
\frac{1}{160}\;B_{\mu \to e\gamma} 
\\
\frac{\Gamma_{\mu\,Ti \to e\,Ti}}{\Gamma_{\mu\,Ti \to{\rm capt}}}
&\simeq &
\frac{1}{240}\;B_{\mu \to e\gamma}\,. 
\end{eqnarray}
Clearly, in this case the decay  $\mu\to e\gamma$ 
would play an utmost important role in searching for LFV. \\ [-2pt]

\begin{figure}[t!]
\centering
  \includegraphics[width=7.6cm,height=4.0cm]{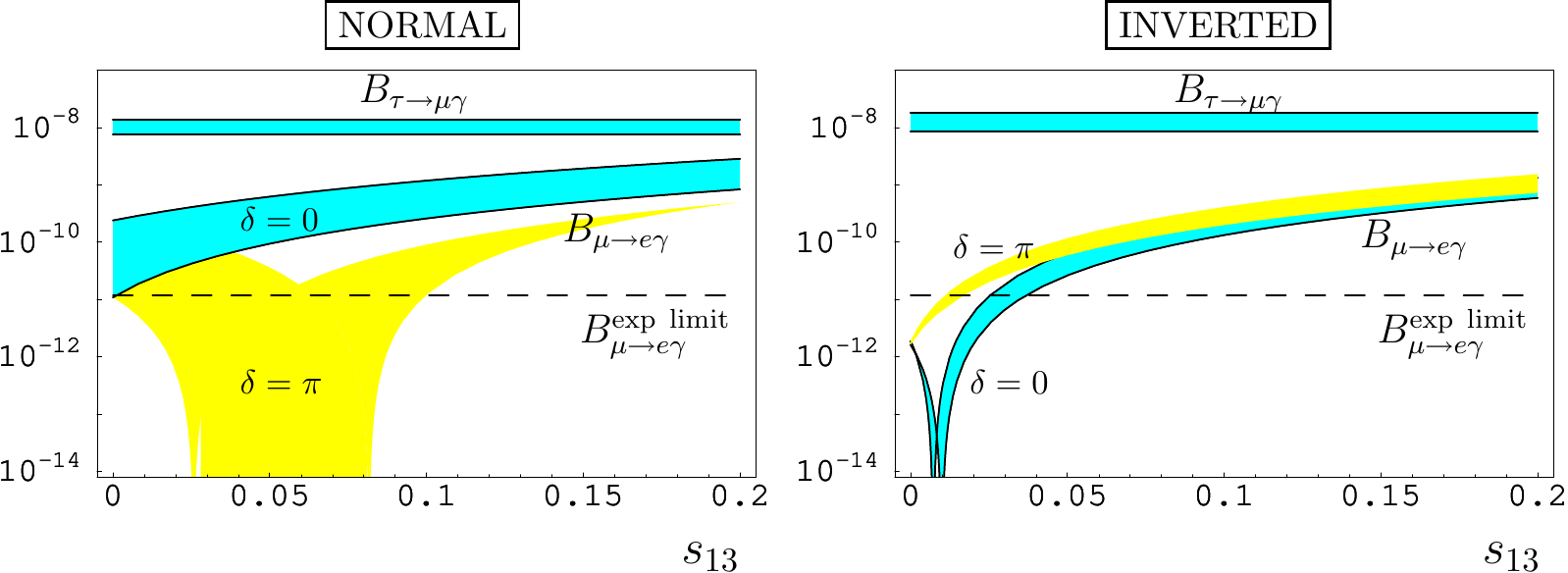}
 \vspace{-0.2cm}
\caption{$B_{\tau \to \mu \gamma}$ and $B_{\mu \to e \gamma}$ as a
  function of $\sin\theta_{13}$ for the CP conserving cases
  $\delta=0,\,\pi$ and $\Lambda_{NP} \sim 10^{-4} (2 v \mu_L)^{1/2}$.
  The shading corresponds to a lightest $\nu$ mass in the range 0 -
  0.02~eV.  (From ref.~\cite{Cirigliano:2005ck}.)
\label{fig:fig5P} 
}
\end{figure}

\noindent
{\it 2. Off-shell photonic and contact operators with quarks}.\\[2pt]
They can give important contributions in particular to $\mu-e$
conversion in atoms, and have the form:
\begin{eqnarray}
  \nonumber
  O^{(H)}_{LL} &=& \bar \ell_\alpha\gamma^\mu\tau^a 
 \Delta_{8}^{\alpha\beta}
\, \ell_\beta\cdot\left( H^\dagger \tau^aiD_\mu H\right)\,,  \\
  \nonumber
O^{(Q)}_{LL} &=& \bar \ell_\alpha\gamma^\mu \tau^a
 \Delta_{8}^{\alpha\beta}
\, \ell_\beta\cdot \left(\bar Q_L\tau^a\gamma_\mu Q_L \right)\,,\\
O^{(q)}_{LL} &=& \bar \ell_\alpha\gamma^\mu  
\Delta_{8}^{\alpha\beta}
\, \ell_\beta\cdot \left(\bar q_R\,\gamma_\mu\, q_R \right)\,,
\end{eqnarray}
where $\tau^a=(1,\vec\tau)$ with $\vec\tau$ the $SU(2)$ matrices 
and $q_R=u_R,\,d_R$ denotes  the RH quarks.  \\ [-2pt]

\noindent
{\it 3. Four leptons contact operators}.\\[2pt]
They can be particularly relevant for $\ell \to 3\ell'$ decays.  
The leading operators have the form:
\begin{eqnarray}
\nonumber
O^{(4\ell)}_{LL} 
& =&\bar \ell_\alpha\gamma^\mu  \tau^a
 \Delta_{8}^{\alpha\beta}
\ell_\beta\cdot\left( \bar \ell_L\tau^a\gamma_\mu \ell_L\right) \\
O^{(2\ell\,2e)}_{LL} & =&\bar \ell_\alpha\gamma^\mu 
 \Delta_{8}^{\alpha\beta}
\ell_\beta\cdot\left( \bar e_R\,\gamma_\mu,\ e_R\right)\,.  
\end{eqnarray}

Clearly, in the general case when operators of type 2. and 3.  are not
particularly suppressed with respect to the operators in 1., searches
for $\mu-e$ conversion in nuclei and for LFV decays like $\mu \to 3e$,
$\tau\to 3\mu$ etc. become equally important than $\mu\to e\gamma$ to
search for LFV.  Here we only consider the radiative decays $\ell\to
\ell'\gamma$, but a detailed analysis of many others LFV processes
within the first MLFV scenario can be found
in~\cite{Cirigliano:2006su}.

\begin{figure}[t!]
  \centering
  \includegraphics[width=3.8cm,height=4.0cm]{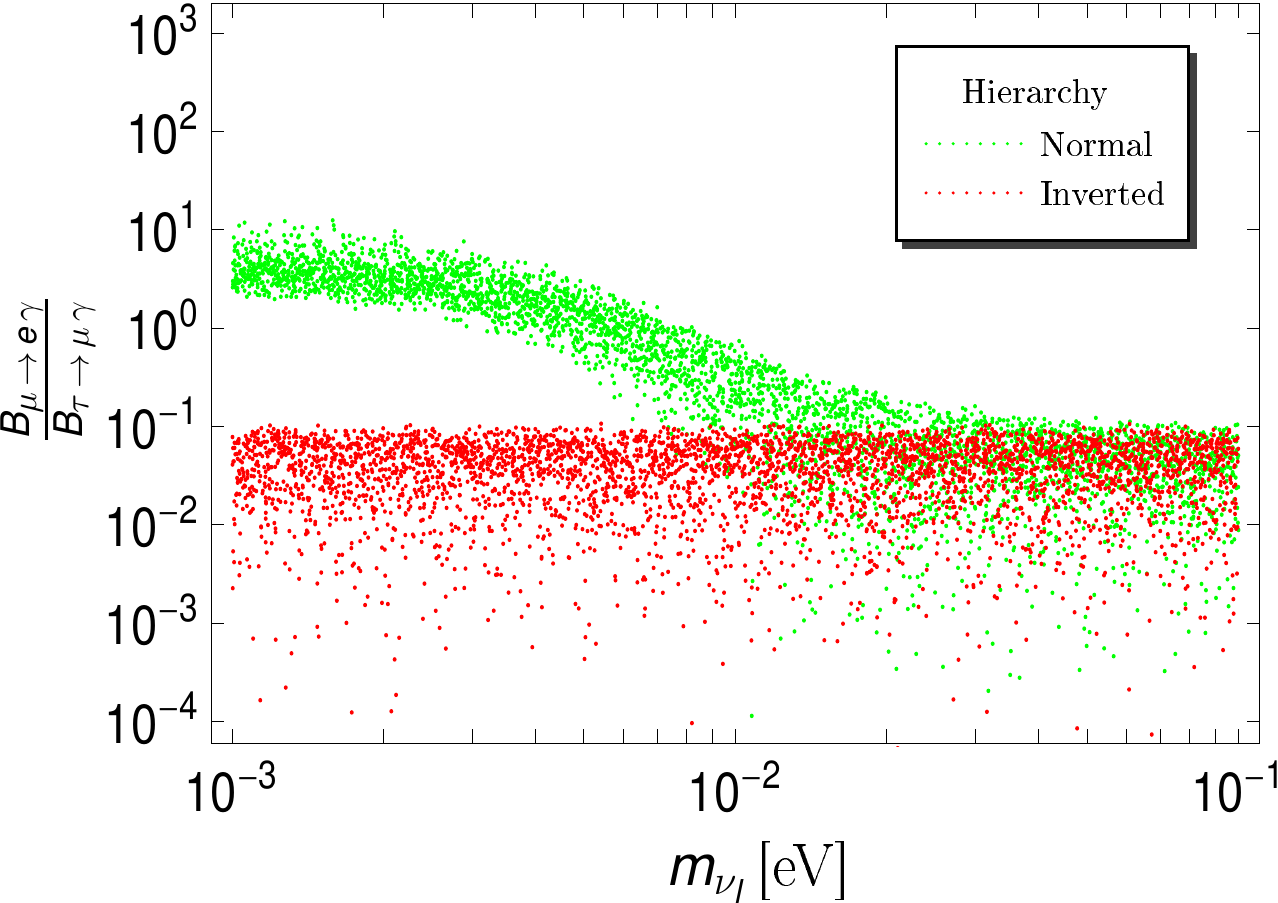}
 \includegraphics[width=3.8cm,height=4.0cm]{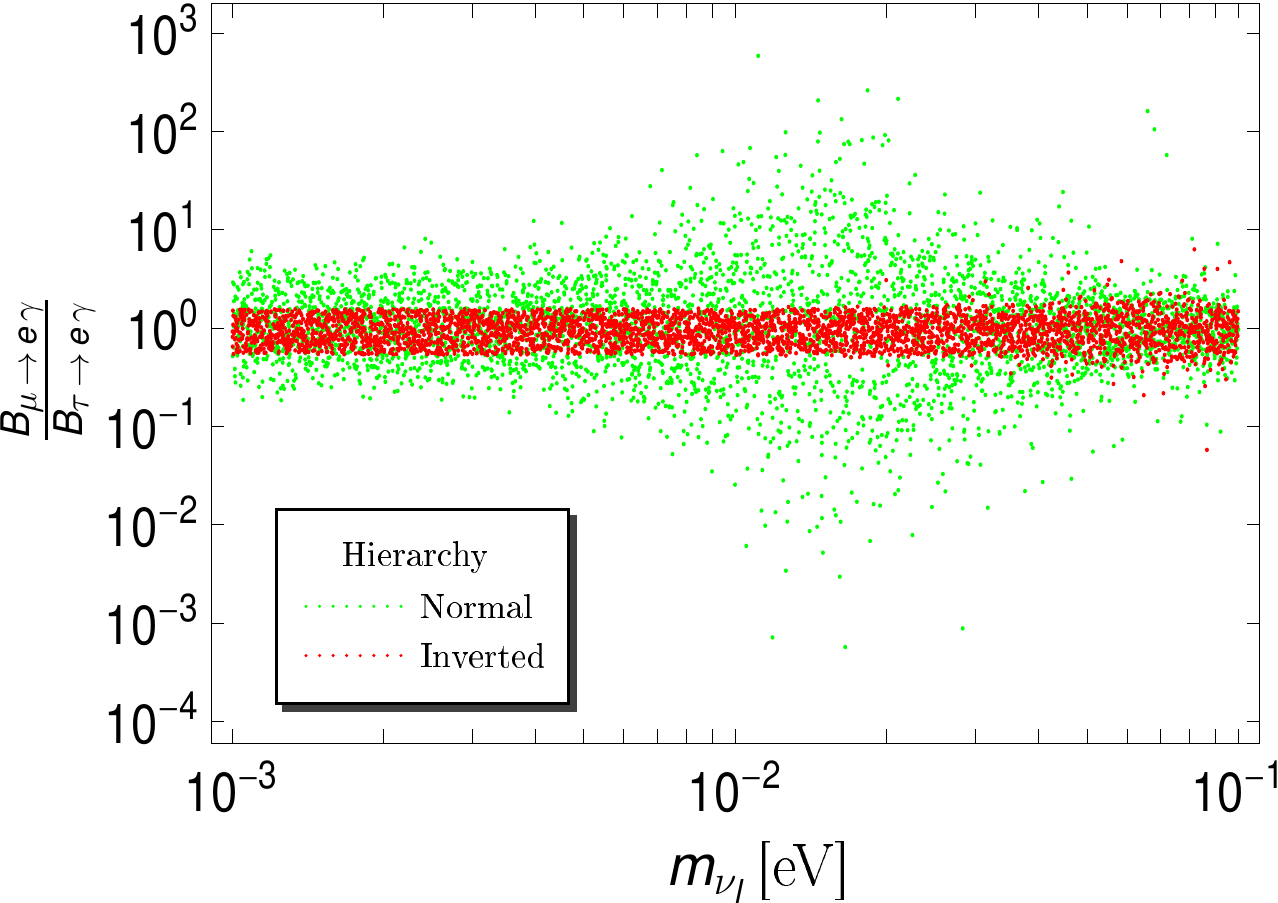}
\vspace{-0.2cm}
\caption{The ratios
  $\frac{B_{\mu\,\rightarrow\,e\,\gamma}}{B_{\tau\,\rightarrow\,\mu\,\gamma}}$
  (left) and
  $\frac{B_{\mu\,\rightarrow\,e\,\gamma}}{B_{\tau\,\rightarrow\,e\,\gamma}}$
  (right) as a function of the lightest neutrino mass.  
Green-lighter points correspond to normal hierarchy, red-darker points
  to inverted hierarchy.
(From ref.~\cite{Alonso:2011jd}.)
\label{fig:brm1}}
\vspace{-.3cm}
\end{figure}

\subsection{Phenomenology}
Let us now discuss for the two cases at hand, the dependence of LFV
processes on low energy parameters. We concentrate on the radiative
decay $\ell_i\to \ell_j\,\gamma$ and on the effects of on-shell
photonic operators $O^{(F)}_{RL}$.  The relevant LFV structure is
$\Delta_8$ defined respectively in eqs.~(\ref{eq:ON}) and
(\ref{eq:MMLFV2}), we also assume that all $c_i$ in \eqn{eq:LowE} are of
$\cO(1)$.  We compare the relevance of different decay channels by
means of the normalized branching fractions:
\begin{equation}
\label{eq:Br}
 B_{\ell_i\to \ell_j\gamma} \equiv  
\frac{\Gamma_{\ell_i\to \ell_j\gamma}}{\Gamma_{\ell_i\to \ell_j\nu_i\bar{\nu}_j}}.
\end{equation}
When the flavour symmetry $O(3)_N\times CP$ is assumed, one observes
the pattern $B_{\tau \to \mu \gamma} \gg B_{\mu\to e \gamma}$ ($\sim
B_{\tau \to e \gamma} $), which is a consequence of the suppression of
LFV effects when the lightest neutrinos mass eigenvalues are involved.
This is illustrated in Fig.~\ref{fig:fig5P}, taken from
ref.~\cite{Cirigliano:2005ck} that depicts the normalized branching
fractions $B_{\tau \to \mu \gamma}$ and $B_{\mu\to e \gamma}$ assuming
a NP scale $\Lambda_{NP} \sim 10^{-4} \sqrt{2 v \mu_L}$.  For a given
choice of $\delta=0$ or $\pi$ (corresponding to CP conservation), the
strength of the $\mu\to e$ suppression is very sensitive to whether
the hierarchy is normal (NH) or inverted (IH).  For $\delta = 0$ the present
experimental limit on $B_{\mu \to e \gamma}$ allows large values of
$B_{\tau \to \mu \gamma}$ only for the IH, whereas for
$\delta = \pi$, a large region with a sizable $B_{\tau \to \mu
  \gamma}$ is allowed only for the NH.  Note that the
overall vertical scale in this figure depends on both the ratio $(v
\mu_L) /\Lambda_{NP}^2$ and on the value of the lightest neutrino mass,
and that a large hierarchy $\Lambda_{NP}/\mu_L \ll 1$ is required to
obtain observable effects.

When the assumed flavour symmetry is $SU(3)_{\ell+N}$, the main
distinctive feature with respect to the previous case is that, due to
the inverse $\mathbf{m}_\nu$ dependence in \eqn{eq:MMLFV2}, LFV
processes are {\it enhanced} when the lighter neutrinos masses are
involved. This implies, in particular, a potentially strong
enhancement of $\mu\to e\gamma$ in the (NH) case. This is better
highlighted by studying ratios of branching ratios for different decay
channels, since they simply reduce to ratios of the modulus squared of
the corresponding $\Delta_8$ entries:
\begin{equation}
\frac{B_{\ell_i\,\rightarrow\,\ell_j\,\gamma}}
{B_{\ell_k\,\rightarrow\,\ell_m\,\gamma}}=
\frac{\left|\left( \Delta_8\right)_{ij}\right|^2}
{\left|\left(\Delta_8\right)_{km}\right|^2}\,.   
\end{equation}
Figure~\ref{fig:brm1} (taken from ref.\cite{Alonso:2011jd}) shows two
scatter plots generated with random values for the quantities
$\Delta_8 \sim U\,\frac{1}{\mathbf{m}_\nu^2}\,U^\dagger$, obtained by
allowing the neutrino parameters to vary within their (approximate)
2$\sigma$ c.l. experimental intervals~\cite{nudata}.  In the left
panel we plot, as a function of the lightest mass eigenvalue, the ratio
$\frac{B_{\mu\,\rightarrow\,e\,\gamma}}{B_{\tau\,\rightarrow\,\mu\,\gamma}}$,
and in the right panel the ratio
$\frac{B_{\mu\,\rightarrow\,e\,\gamma}}{B_{\tau\,\rightarrow\,e\,\gamma}}$.
Results for the NH ($m_{\nu_l}=m_{\nu_1}$) correspond to the
green-lighter points, while the IH ($m_{\nu_l}=m_{\nu_3}$) to the
red-darker points.  From the first panel we see that for NH and small
values of $m_{\nu_1} \lesssim 10^{-2}\,$eV we generically have
$B_{\mu\,\rightarrow\,e\,\gamma}\,>\,B_{\tau\,\rightarrow\,\mu\,\gamma}$.
The enhancement of $B_{\mu\,\rightarrow\,e\,\gamma}$ is obviously due
to ${{\bf m}^2_\nu}$ appearing in the denominator of $\Delta_8$, and
can be of a factor of a few. In the limit of $m_{\nu_1}\ll
m_{\nu_{2,3}}$, and using the best fit values of the mixing angles, we
have:
$\frac{B_{\mu\,\rightarrow\,e\,\gamma}}{B_{\tau\,\rightarrow\,\mu\,\gamma}}
 \approx  7.3\; (3.2) $ for $\delta=0$ ($\delta=\pi$).
When $m^2_{\nu_1}\gg \Delta\,m^2_{\rm sol}$ and $m_{\nu_1} \approx
m_{\nu_2}$, the contributions to $\mu\,\rightarrow\,e\,\gamma$
proportional to $\theta_{12}$ suffer a strong GIM suppression, and the
decay rate becomes proportional to $\theta_{13}^2$ .  This behavior is
seen clearly in Fig.~\ref{fig:brm1} (left) for values of $m_{\nu_1}
\approx 10^{-2}\,$eV.  For IH, in the limit $m_{\nu_3}\ll
m_{\nu_{1,2}}$ and independently of the value of $\delta$ we obtain:
$\frac{B_{\mu\,\rightarrow\,e\,\gamma}}{B_{\tau\,\rightarrow\,\mu\,\gamma}}
\approx 2\, s_{13}^2$.
Approximately the same  result is obtained also 
in the limit of large masses $m_{\nu_i} \gg \sqrt{\Delta m^2_{\rm atm}}$,  
which explains why for $m_{\nu_1}\to 10^{-1}\,$eV the results for  IH and NH 
converge.

Results for the ratio of the $\mu$ and $\tau$ radiative decays into
electrons are depicted in the right panel in Fig.~\ref{fig:brm1}. At a
glance we see that for both NH and IH the $\mu/\tau$ ratios for decays
into electrons remain centered around one for all values of
$m_{\nu_l}$.  Needless to say, since the ratio of normalized branching
ratios of other LFV processes like for example $B_{\mu\to 3e}$,
$B_{\tau\to 3\mu}$, $B_{\tau\to 3e}$ are controlled by the same LFV
factors $\Delta_8$, they are characterized by a completely similar
pattern of enhancements/suppressions.

In view of the ongoing high sensitivity searches for LFV
processes~\cite{MEG}, besides comparing the rates for different LFV
channels, an estimate of the absolute values of the branching
fractions is also of primary interest.  In the most favorable case, in
which $\Delta_8$ is a matrix with $\cO(1)$ entries, a rough estimate
gives:
\begin{equation}
  \label{eq:abs}
B_{\mu\,\rightarrow\,e\,\gamma} \approx 1536\,\pi^3\, \alpha\,
\frac{v^4}{\Lambda_{NP}^4 }\,.  
\end{equation}
When compared with the experimental limit $ B^{\rm
  exp}_{\mu\,\rightarrow\,e\,\gamma}< 10^{-11}$~\cite{MEGA} this
allows us to conclude that the scale of NP should be rather large:
$\Lambda_{NP} \gtrsim 400\,$TeV.

In summary, this second MLFV scenario~\cite{Alonso:2011jd} is
characterized by a quite different phenomenology from the first one
since it allows the branching fraction
$B_{\mu\,\rightarrow\,e\,\gamma}$ to dominate over
$B_{\tau\,\rightarrow\,\mu\,\gamma}$ and
$B_{\tau\,\rightarrow\,e\,\gamma}$.  The enhancement with respect
$B_{\tau\,\rightarrow\,\mu\,\gamma}$ that occurs in the NH case does
not exceed a factor of a few, but it is parametric in the small values
of $m_{\nu_1}$.  The strong enhancement with respect to
$B_{\tau\,\rightarrow\,e\,\gamma}$ instead is due to accidental
cancellations that suppress this process, and that become particularly
efficient when $\delta$ is close to zero.

\section*{Acknowledgments}
I thank the authors of ref.~\cite{Cirigliano:2005ck} for permission  
to include fig.~\ref{fig:fig5P} in this review.

 

 
 

\end{document}